\documentclass[aps,prb,showpacs,groupedaddress,twocolumn,showkeys]{revtex4}
\usepackage{graphicx}
\usepackage{amssymb}
\usepackage{verbatim}
\begin{document}
\setlength{\unitlength}{1cm}
\title{Electron conductivity and second generation Composite Fermions}
\author{Matteo Merlo$^{1}$, Nicodemo Magnoli$^{2}$, Maura Sassetti$^{1}$
and Bernhard Kramer$^{3}$} \affiliation{
  $^{1}$Dipartimento di Fisica, INFM-LAMIA, 
  Universit\`{a} di Genova, Via
  Dodecaneso 33, I-16146 Genova, Italy \\
  $^{2}$Dipartimento di Fisica, INFN,
  Universit\`{a} di Genova, Via
  Dodecaneso 33, I-16146 Genova, Italy \\
  $^{3}$I. Institut f\"ur Theoretische Physik,
  Universit\"at Hamburg,
  Jungiusstra\ss{}e 9, D-20355 Hamburg, Germany }
\date{\today}
\begin{abstract}
  The relation between the conductivity tensors of Composite Fermions and
  electrons is extended to second generation Composite Fermions. It is shown
  that it crucially depends on the coupling matrix for the Chern-Simons gauge
  field. The results are applied to a model of interacting Composite Fermions
  that can explain both the anomalous plateaus in spin polarization and the
  corresponding maxima in the resistivity observed in recent transport
  experiments.
\end{abstract}
\pacs{71.10.Pm; 73.43.Cd; 73.43.Nq}
\keywords{composite fermions; conductivity}
\maketitle

Many aspects of the Fractional Quantum Hall Effect (FQHE)\cite{stormer82} can be 
understood in terms of the Composite Fermion (CF) model.\cite{Jain89,LF91,HLR93} In 
the field theoretical approach, CF are constructed by attaching an even number 
$\tilde\varphi$ of flux quanta to an electron via a Chern-Simons (CS) gauge field. 
In mean field approximation, the Coulomb repulsion between the electrons is removed 
and the gauge field partly cancels the external vector potential, so that CF 
experience a reduced effective magnetic field $\Delta B$ and are otherwise free. For 
certain values of the electron filling factor $\nu$, $\Delta B=0$ and CF form a 
Fermi sea. For $\Delta B\neq 0$, CF develop Landau levels. Thus, the principal 
sequence of FQHE states of electrons at filling factors $\nu=1/3,2/5,\ldots$ can be 
viewed as the QHE of non-interacting CF at integer fillings $p=1,2,\ldots$. This 
explains many of the observed FQHE states. Fluctuations of the gauge field around 
the mean value induce an effective interaction between the CF. In the simplest 
approximation, this can be incorporated into a finite effective mass of the CF, 
$m^*\propto\sqrt{B}$.\cite{HLR93} The quantization of the conductivity of the 
electrons to the correct fractional values of $e^2/h$ is obtained by including in 
Random Phase Approximation (RPA) the gauge field fluctuations.\cite{simon}


Including the electron spin,\cite{halperin} the mean field CF picture can be used to 
understand the spin polarization of the ground state.\cite{spintrans} This has been 
detected in experiments at fixed $\nu$, where the spin polarization $\gamma$ has 
been measured optically as a function of the magnetic field. It shows crossovers 
between non-fully polarized states at low magnetic field and completely polarized 
states at high magnetic field, $\gamma=1$.\cite{Kukushkin} In these experiments, 
plateaus of intermediate polarization have been detected. These cannot be explained 
by using the mean field approach. It has been argued that the residual CF 
interaction can be attractive, thus leading to an instability towards pairing in 
spin singlet states that can account for the partly polarized plateaus.\cite{psss}   
There are two possibilities, namely particle-particle pairs in analogy to the Cooper 
pairs in superconductivity, and particle-hole pairs. Chern-Simons theories do indeed 
play a role in suggesting novel possibilities for ground states, even if  
quantitative estimates of such suggestions should be subjected to independent 
testing, e.g. in the form of variational wavefunction calculations.

A doublet Chern-Simons gauge field has been proposed to include spin.\cite{mr96} 
Obviously, there are several possibilities for flux attachments in this case since 
different CS fluxes can be associated to electrons with opposite spins. The 
Lagrangian density for the system of interacting electrons in a magnetic field is 
${\cal L} = {\cal L}_{\rm F}  + {\cal L}_{{\rm CS}}  + {\cal L}_{{\rm I}} + {\cal 
L}_{{\rm Z}}$ where ($\hbar=c=1$ for the moment)
\begin{eqnarray} \label{eq:lagr}
{\cal L}_{\rm F}({\bf r},t)&\!=\!&\sum_{s=\uparrow,\downarrow}
\psi^{\dagger}_{s}({\bf r},t)
\Big\{i\partial_{t}+\mu+e a_0^{s}({\bf r},t)\nonumber \\
& & \hspace{-1cm}
-\frac{1}{2m_{}}\Big[i\nabla+e \Big({\bf A}({\bf r})-{\bf a}^{s}({\bf
r},t)\Big)\Big]^2 \Big\}\psi_{s}({\bf r},t), \\
{\cal L}_{{\rm CS}}({\bf r},t) &\!=\!&-\frac{e}{\phi_{0}}
\sum_{s,s'} \Theta_{ss'} \,a_0^{s}({\bf r},t)\, {\bf
\hat{z}}\cdot\nabla\times{\bf a}^{s'} ({\bf r},t),\\
{\cal L}_{{\rm I}}({\bf r},t)&\!=\!&-{\frac{1}{2}}\sum_{s,s'} \int \!{\rm d}{\bf r}'
n_{s}({\bf r},t) V({\bf r}-{\bf r}') n_{s'} ({\bf r}',t).
\end{eqnarray}
Here, ${\cal L}_{\rm F}$ represents the kinetic energy of the CF ($m$ electron
band mass, $\mu$ chemical potential), ${\cal L}_{{\rm CS}}$ the Chern-Simons
doublet field ($\phi_{0}=hc/e$ flux quantum, ${\bf \hat{z}}$ unit vector
perpendicular to the 2D system), with matrix-valued coupling strength
$\Theta_{ss'}$, ${\cal L}_{{\rm I}}$ the interaction, and ${\cal L}_{{\rm Z}}$
the Zeeman energy.  Furthermore, $n_{s}({\bf r},t)\equiv
\psi^{\dagger}_{s}({\bf r},t) \psi_{s}({\bf r},t)$ is the density of the
Fermions with spin orientation $s$, ${\bf A}$ the vector potential of the
external magnetic field, $(a_{0},{\bf a})$ the CS gauge field, and $V({\bf
  r})=e^2/\epsilon r$ the Coulomb interaction potential ($\epsilon$ host
material dielectric constant).

The attachment of flux quanta is achieved by the CS-term as can be
seen by minimizing ${\cal L}$ with respect to $a_{0}^{s}$,
\begin{equation}
\label{eq:constraint} \sum_{s'}\Theta_{ss'} {\bf \hat{z}}\cdot\nabla\times{\bf 
a}^{s'} ({\bf r},t) = \phi_0 n_s({\bf r},t)
\end{equation}
with the symmetric $2\times2$  coupling matrix
\begin{eqnarray}
  \label{eq:kdd}
\Theta=\left(
  \begin{array}{cc}
\theta_{1}&\theta_{2}\\
\theta_{2}&\theta_{1}
  \end{array}\right).
\end{eqnarray}
A given CF ''sees'' a CF with like$\backslash$unlike spin as carrying a number
$\theta_{1\backslash 2}/(\theta_1^2-\theta_2^2)$ of flux quanta.

Different choices for $\Theta$ map to CF systems with different properties. This is 
obvious in the diagonal basis ${\bf a}^\pm=1/2({\bf a}^\uparrow \pm {\bf
  a}^\downarrow)$, in which Eq.~(\ref{eq:constraint}) reads
\begin{eqnarray}\label{eq:cos}
\phi_0 [n_\uparrow({\bf r},t)+n_\downarrow({\bf r},t)]
&=&\theta_+ \, \hat{z}\cdot \nabla\times {\bf a}^+ \nonumber \\
\phi_0[n_\uparrow({\bf r},t)-n_\downarrow({\bf r},t)]
&=&\theta_-\, \hat{z}\cdot \nabla\times {\bf a}^- ,
\end{eqnarray}
with $\theta_{\pm}=2(\theta_1 \pm \theta_2)$.  For instance, the singular choice
$\theta_-=0$ demands that \emph{locally} $n_\uparrow({\bf
  r},t)=n_\downarrow({\bf r},t)$ and therefore describes a spinless boson of
charge $2e$.
 
The attachment of the same flux to all electrons independently of the spin can be 
obtained with $\theta_-\to\infty$ and $\theta_+$ finite. This decouples ${\bf a}^-$ 
and leads to a single CS-field ${\bf a}^+$ coupled to all CF. This is the 
conventional method for describing partly polarized states.\cite{mr96,lf95} It maps 
electrons with spin $s$ at filling factor $\nu$ to composite particles consisting of 
one electron with spin $s$ and $\tilde\varphi=\theta_+^{-1}$ flux quanta in an 
effective field $\Delta B^s=\Delta B=B-n(\mathbf{r})\tilde\varphi \phi_0$ that is 
independent of the spin, with $n(\mathbf{r},t)=n_\uparrow({\bf 
r},t)+n_\downarrow({\bf r},t)$ the total density and $\tilde\varphi$ even to ensure 
fermionic statistics.

At mean field level, $n(\mathbf{r})$ is replaced by its average $n_e$,
$\Delta B$ is uniform and induces quantization into CF Landau levels
(CFLL). One can define CF filling factors $p_s=\phi_0 n_s/\Delta B$ that can
be related to $\nu$. This gives the principal sequence
$\nu={(p_\uparrow+p_\downarrow})/
{[\tilde\varphi (p_\uparrow+p_\downarrow)\pm 1]}$
with $p_\uparrow$, $p_\downarrow$ integer numbers of filled spin up/down CFLL.
The limit $\Delta B=0$ corresponds to a Fermi sea at $\nu=1/\tilde\varphi$ and
is obtained for $p=p_\uparrow+p_\downarrow \to\infty$.

The cyclotron energies $\hbar\Delta\omega_c=\hbar e \Delta B/m c$ for both
spins scale incorrectly with the magnetic field at fixed filling $p$.
Dimensional analysis\cite{HLR93} and explicit calculations\cite{morf,stern}
yield a renormalized \mbox{$\hbar\Delta\omega_c^* \propto E_{\rm C}/(\tilde\varphi p \pm 1)$},
with the Coulomb energy $E_{\rm C}= e^2/\epsilon \ell$ and $2\pi B
\ell^2=\phi_0$ defines the magnetic length $\ell$. This can be recovered in
mean field approximation by replacing $m$ with an effective mass
\mbox{$m^*=\alpha m_0 \sqrt{B[\textrm{Tesla}]}$}, with a sample dependent
parameter $\alpha$ and $m_0$ the electron mass.\cite{park} For
$\tilde\varphi=2$, relevant for filling factors $1/3<\nu<1$, one finds
$\hbar\Delta\omega_c^*={\hbar e\sqrt{B}}/[{m_0 c \alpha(2p\pm
  1)}]$.

Combining the Landau quantization with the Zeeman term yields spin-split CF
Landau levels ($\mu_{\rm B}=\hbar e/2m_0c$)
\begin{equation}\label{eq:energy}
 E_{nps}(B)=\left(n+\frac{1}{2}\right)
\hbar\Delta \omega_c^* +\frac{1}{2}sg\mu_{\rm B} B,
\end{equation}
with $s=\uparrow,\downarrow=\pm 1$ and
the CF Land\'e factor  $g$.

\begin{figure}[h!]
  \begin{center}
  \includegraphics[scale=0.83]{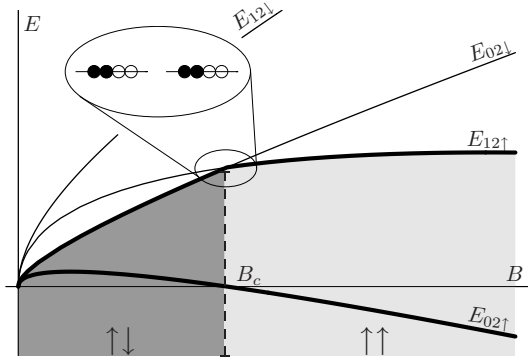}
  \caption{CFLL for $p=2$
    ($\nu={2}/{5}$). The four lowest $E_{nps}$ of Eq.~(\ref{eq:energy}) are
    shown with arbitrary scales for $E$ and $B$. Bold: lowest $p$ filled CFLL
    that constitute the ground state; dark grey: $p_\uparrow=p_\downarrow=1$,
    $\gamma=0$; light grey: $p_\uparrow=2, p_\downarrow=0$, $\gamma=1$. At the
    crossover field $B_c$, two levels intersect.}
           \label{fig:levels}
  \end{center}
\end{figure}

Various properties of the FQHE states at $\nu=p/(2p\pm1)$ can be determined
using this spectrum and the ground state obtained by filling the lowest $p$
CFLL. For instance, the main crossovers in the spin polarization
\begin{equation}
\gamma =\frac{n_\uparrow-n_\downarrow}{n_\uparrow+n_\downarrow}
=\frac{p_\uparrow-p_\downarrow}{p_\uparrow+p_\downarrow}
\end{equation}
at fixed $\nu$ can be associated with intersections between CFLL
with opposite spins at the Fermi energy (Fig.\ref{fig:levels}).

The parameter $\alpha$ and the CF $g$-factor have been determined by
fitting Eq.~(\ref{eq:energy}) to measurements of the crossover field $B_{\rm
  c}$\cite{spintrans} and the activation gap.\cite{prlmariani} This yields a
step-like polarization. However, the experimentally detected
half-polarized plateaus cannot be reproduced.

Close to $B_c$ (Fig.\ref{fig:levels}), $E_{02\downarrow}$ and $E_{12\uparrow}$
are almost degenerate and the residual CF interaction is expected to become
relevant.\cite{murthy} We have studied this case by assuming that exactly at
degeneracy both levels are half-filled. We have performed another Chern-Simons
transformation to create second-generation CF ($^2$CF). The lowest CFLL
$E_{02\uparrow}$ is not involved in the level crossing. As before, a suitable
choice of the coupling matrix $\Theta$ is necessary to specify the
transformation. In this case, it appears natural to perform the flux
attachment in an independent way, namely assuming diagonal coupling
$\Theta=\textrm{diag} (\theta)$. While this choice is perfectly legitimate, it
differs from the first CS transformation in that both gauge fields ${\bf
  a}^\pm$ survive and have to be considered separately.  This will turn to be
crucial when evaluating the conductivity.

The degenerate system at $B_c$ is described by a Lagrangian equivalent to the one 
above but without the Zeeman term. The constraint Eq.~(\ref{eq:constraint}) is now 
$\theta \,{\bf\hat{z}}\cdot\nabla\times{\bf a}^{s} ({\bf r},t) = \phi_0 n_s({\bf 
r},t)$ where in mean field approximation $n_\uparrow=n_\downarrow$ because of the 
assumption of half-filled levels.  Each species of CF therefore couples separately 
to the same-spin CS field. The choice $\theta=1/2$ preserves fermionic statistics 
and maps the system on two Fermi seas at $B=0$ that are coupled by the interaction.

An effective interaction between $^2$CF
\begin{eqnarray}
&V^{s,s'}({\bf k},{\bf k'},{\bf q};\Omega)
=v^{s}({\bf k},{\bf q})\,v^{s'}({\bf
k'},-{\bf q})\nonumber\\
&\qquad\qquad\times[{\cal D}^{+}({\bf q},\Omega)+
(2\delta_{ss'}-1){\cal D}^{-}({\bf q},\Omega)]
\label{eq:effint}
\end{eqnarray}
is then obtained considering scattering of states with spin $s$, $s'$ and momenta 
${\bf k}$, and ${\bf k'}$ into states with the same spins and momenta $({\bf k}+{\bf 
q})$, $({\bf k'}-{\bf q})$, mediated by the exchange of a gauge field fluctuation 
with momentum ${\bf q}$ and frequency $\Omega$. In Eq.~(\ref{eq:effint}), ${\cal 
D}^{\pm}$ are RPA gauge field propagators associated with the symmetric and 
antisymmetric combinations of spin and $v^s$ are interaction vertices between $^2$CF 
and the CS-field fluctuations. The potential $V^{s,s'}$ is attractive in the 
$s$-wave channel both for particle-particle (p-p) and particle-hole (p-h) pairs with 
zero total spin. New collective ground states $\vert \Phi_{pp}\rangle$, $\vert 
\Phi_{ph}\rangle$ form, characterized by pair-breaking gaps\cite{note}
\begin{eqnarray}\label{eq:gs}
\Delta_{pp} \propto \langle \Phi_{pp} \vert & \psi^\dag_\uparrow \psi^\dag_\downarrow & \vert \Phi_{pp} \rangle \\
\Delta_{ph} \propto \langle \Phi_{ph} \vert & \psi^\dag_\uparrow \psi_\downarrow & \vert \Phi_{ph} \rangle.
\end{eqnarray}
The energies of the paired ground states depend differently on the strength and 
range of the interaction so that crossovers between the p-p and the p-h phases can 
occur.\cite{merlo} Both states are rigid zero-po\-larization states. Flipping a spin 
requires a fini\-te amount of energy given by $\Delta_{pp}$ or $\Delta_{ph}$, both 
of the order $0.5$ K for typical experimental conditions. This can explain the 
half-polarized plateaus observed experimentally\cite{Kukushkin} as signature of the 
condensation into a paired singlet state which does not contribute towards spin 
polarization. The condition for this is that the energy gain by forming the paired 
ground state should be larger than the energy required for exciting the CF to the 
higher level. 
In the example
of Fig.~\ref{fig:levels}, the crossing levels $E_{02\downarrow}$ and
$E_{12\uparrow}$ contribute spin 0, while $E_{02\uparrow}$ has all spins up.
Since the degeneracy is the same for all levels, the total polarization is
$\gamma=1/2$. This can be generalized to other filling factors, such as $3/7$
or $4/9$, where more intersections occur, and one can systematically reproduce
all observed partially polarized plateaus.

The p-p and p-h pairs binding energy are of the same order,
though.  Therefore, it is not possible to distinguish between the two phases
by comparing the predicted plateau widths with experiment.

To find differences in the nature of the paired states, we consider electrical 
transport. Since a CF contains charge and flux, a current $\mathbf{j}$ of CF across 
a closed loop corresponds to a variation of magnetic flux. This yields a CS-electric 
field $\mathbf{E}^{\rm{CS}}$ via Faraday's law 
$\mathbf{E}^{\rm{CS}}=(\tilde\varphi\phi_0/{ec})\hat\mathbf{z}\times\mathbf{j}$ 
where $\hat\mathbf{z}$ is the unit vector perpendicular to the 2D electron system in 
the $(x,y)$-plane. Self-consistency requires that CF respond to both 
$\mathbf{E}^{\rm{CS}}$ and the electric field $\mathbf{E}$,\cite{simon}
\begin{equation}\label{sigma}
\mathbf{j}=\hat\sigma^{\rm CF}(\mathbf{E}+\mathbf{E}^{\rm{CS}})
\end{equation}
with the CF conductivity $\hat\sigma^{\rm CF}$. Defining CS- and  electron
resistivities $\mathbf{E}^{\rm CS}=-\hat\rho^{\rm CS}\mathbf{j}$ and
$\mathbf{E}=\hat\rho\mathbf{j}$, respectively,
\begin{equation}\label{resist}
\hat\rho=\hat\rho^{\rm CF}+\hat\rho^{\rm CS}
\end{equation}
with
\begin{equation}\label{rhocs}
\hat\rho^{\rm CS}= \tilde\varphi \frac{h}{e^2} \left(
\begin{array}{cc}
0 & 1 \\
-1 & 0 \\
\end{array}
\right)
\end{equation}
and $\hat\rho^{\rm CF}=[\hat{\sigma^{\rm CF}}]^{-1}$.  Equation~(\ref{resist})
can be more formally justified using linear response theory in terms of
current-current correlation functions.\cite{HLR93} It has been used
successfully for both FQHE and even-denominator QHE states. Its applicability
depends on the choice of the coupling matrix $\Theta$. It is correct when a
\emph{single} CS-field couples with the \emph{total} density, as for spinless
fermions. It also holds for the above first CS-transformation with the special
choice $\theta_-\to\infty$, finite $\theta_+$, where again a single CS-field,
in this case ${\bf a}^+$, couples to the fermions.\cite{mr98}

Regarding the situation near the crossover points, we need to determine
$\rho_{\rm CF}$. We split the conductivity of the $p$ levels of CF into a
contribution $\hat\sigma_{\rm h}^{\rm CF}$ from the highest level, which
changes and determines the spin transition, and a contribution from the lowest
$p-1$ levels $\hat\sigma^{\rm CF}_{p-1}=\hat\sigma^{\rm CF}- \hat\sigma^{\rm
  CF}_{\rm h}$.  For levels that are not involved in crossings, CF are treated
as non-interacting fermions in an effective magnetic field. Thus,
$\hat\sigma^{\rm CF}_{p-1}$ is the sum of the conductivities $\hat\sigma^{\rm
  LL}$ of the filled Landau levels,
\begin{equation}\label{eq:p-1}
\hat\sigma^{\rm CF}_{p-1}=(p-1)
\hat\sigma^{\rm LL}\equiv (p-1)\frac{e^2}{h}\left(
\begin{array}{cc}
0 & -1 \\
1 & 0 \\
\end{array}
\right).
\end{equation}

The contribution of the highest level depends on whether we are at a spin
transition or not. It is again that of a filled CFLL $\hat\sigma^{\rm CF}_{\rm
  h}=\hat\sigma^{\rm LL}$ for $B\neq B_{c}$. In this case, the resistivity of
quantum Hall states is reproduced,
\begin{eqnarray}\label{eq:hallre}
\rho_{xx}(B)&=&0 \nonumber \\
\rho_{xy}(B) &=&\frac{h}{e^2}\frac{2p+1}{p}=\frac{h}{\nu e^2}.
\end{eqnarray}

At $B\approx B_{c}$, the highest occupied level consists of two crossing
levels, both half-filled, on which a CS-mapping to $^2$CF is done. The
contribution to transport will then depend on the nature of the state formed
at the transition.
It appears natural to consider $\hat\sigma^{\rm CF}_{\rm h}$ using an
expression equivalent to Eq.~(\ref{resist}) but applied to $^{2}$CF and
consistent with our assumptions for $\Theta$. Considering the Lagrangian for
the two transformed intersecting levels, the corresponding current density
will have contributions from both spin up/down $^2$CF according to ${\bf
  j}={\bf j}_\uparrow+{\bf j}_\downarrow$.  Each will induce a fictitious CS
electric field felt only by the fermions with equal spins,
\begin{equation}\label{rhocss}
{\mathbf E}^{\rm CS}_s= -\hat\rho^{\rm CS}\mathbf{j}_s,
\end{equation}
where the CS-resistivity is given by Eq.~(\ref{rhocs}) because it depends on
the number of attached fluxes ($=2$) in both cases. Linear response gives
\begin{eqnarray}
{\bf j}_s&=&\sum_{s'}\hat\sigma_{ss'}^{^2\rm CF}
\Big( {\bf E}+{\bf E}^{\rm CS}_{s'}\Big),
\end{eqnarray}
with $\hat\sigma_{\uparrow\uparrow}^{^2\rm CF}
=\hat\sigma_{\downarrow\downarrow}^{^2\rm CF}$ and
$\hat\sigma_{\uparrow\downarrow}^{^2\rm CF}
=\hat\sigma_{\downarrow\uparrow}^{^2\rm CF}$ because of spin up/down symmetry.
Inserting
Eq.~(\ref{rhocss}),
one gets
\begin{equation}
{\mathbf j}=\hat\sigma_h^{\rm CF} {\bf E}=
2(\hat\sigma_{\uparrow\uparrow}^{^2\rm CF}+
\hat\sigma_{\uparrow\downarrow}^{^2\rm CF}){\bf E}-
(\hat\sigma_{\uparrow\uparrow}^{^2\rm CF}+
\hat\sigma_{\uparrow\downarrow}^{^2\rm CF})\hat\rho^{\rm CS}{\mathbf j},
\end{equation}
so that the equivalent of Eq.~(\ref{resist}) becomes
\begin{equation}\label{newresist}
\hat\rho_h^{\rm CF}=\hat\rho^{^2 \rm CF}+\frac{1}{2}\hat\rho^{\rm CS},
\end{equation}
where $\hat\rho^{^2 \rm CF}=[2(\hat\sigma_{\uparrow\uparrow}^{^2\rm CF}+
\hat\sigma_{\uparrow\downarrow}^{^2\rm CF})]^{-1}$, and $\hat\rho_h^{\rm
  CF}=[\hat\sigma^{\rm CF}_{\rm h}]^{-1}$ is the resistivity of the degenerate
levels of $^{1}$CF.

We now calculate the resistivity of the $^{2}$CF. For non-interacting $^2$CF
in unperturbed Fermi seas, one expects
\begin{equation}\label{eq:dff}
\hat\rho^{^2\rm CF}=\frac{h}{e^2}\left(
\begin{array}{cc}
\rho_0 & 0 \\
0 & \rho_0 \\
\end{array}
\right),
\end{equation}
with $\rho_0$ the resistance due to impurity scattering.

The p-h pairing leads to a modification of the Fermi surfaces that leaves
unchanged the response of the system in the presence of impurities.  The
resistivity $\hat\rho_{ph}^{^2{\rm CF}}$ is given by Eq.~(\ref{eq:dff}).  On
the other hand, since p-p pairing leads to an instability of the CF Fermi
sphere in analogy to the BCS theory of superconductivity, the diagonal
elements of the resistivity vanish, even with disorder.

Using Eq.~(\ref{newresist}) we then have
\begin{eqnarray}
 \hat\sigma^{\rm CF}_{\rm h}|_{ph}
&=&\frac{e^2}{h}\frac{1}{\rho_0^2+1}\left(
\begin{array}{cc}
\rho_0 & -1 \\
1 & \rho_0 \\
\end{array}
\right) \nonumber\\
\hat\sigma^{\rm CF}_{\rm h}|_{pp}&=&\frac{e^2}{h}\left(
\begin{array}{cc}
0 & -1 \\
1 & 0 \\
\end{array}
\right).
\end{eqnarray}

Inserting $\hat\sigma^{\rm CF}_{\rm h}|_{ph}$ into (\ref{resist}), one finds
for the p-h phase
\begin{eqnarray}\label{rhokk}
\rho_{xx}(B_c)&=&\frac{h}{e^2} \,\frac{\rho_0}{p^2 +
(p -1)^2\,\rho_0^2}
 \nonumber \\
\rho_{xy}(B_c)&=&\frac{h}{e^2} \left[ \frac{2p+1}{p}+ \frac{\rho_0^2(p-1)}{p^3+p(p-1)^2\rho_0^2}\right].
\end{eqnarray}
The result for the p-p state is obtained with $\rho_0= 0$ which
reproduces Eq.~(\ref{eq:hallre}). Thus, the two states can be distinguished
measuring the resistance. For the p-h state $\rho_{xx}$ and
$\rho_{xy}$ will deviate from the ideal quantum Hall values.

Recently, non-universal behavior in $\rho_{xx}$ and $\rho_{xy}$ has been observed\cite{smettt}
near the spin transition, accompanied by
hysteresis induced by the coupling between the electron system and the nuclear
spins.
Previously, this was attributed to ferromagnetic domains in the quantum Hall
system which can produce deviations in the resistance components from their
ideal values due to scattering at the domain walls. If the above p-h state was
formed, this could also account for both the half-polarized spin plateaus \emph{and}
the anomalous behavior of the resistance.

In conclusion, we have extended the standard procedure to relate the conductivities 
of CF and electrons to include multiple flux attachment. This is important for 
understanding the behavior of quantum Hall systems at filling factors where 
transitions between ground states with different spins occur. In the presence of 
interactions between CF, we have found that for p-h pairing at the spin transition 
the resistance indicates a breakdown of the quantum Hall phase. If p-p (Cooper) 
pairs are formed the resistance is not affected. 
Finally, we would like to mention that experiments suggest similar features to occur
 at other fillings. For instance, if three-quarters/one-quarter filling of CF spin-up/down 
levels, respectively, was forming a stable correlated many body state, this could 
account for the three-quarter polarized state observed recently at electron filling 
$\nu = 2/3$.\cite{fetal01} 

 {\bf Acknowledgments:} Financial 
support from the EU via MCRTN-CT2003-504574 and from the DFG via Project Kr 627/12 
of the special research program ''Quanten-Hall-Effekt'' is gratefully acknowledged.

\end{document}